\documentclass[%
 aip,
 jmp,%
 amsmath,amssymb,
 reprint,%
]{revtex4-1}

\usepackage{graphicx}
\usepackage{dcolumn}
\usepackage{bm}
\usepackage{color}
\definecolor{darkgreen}{rgb}{0.1,.6,.1}
\definecolor{greenblue}{rgb}{0.0,.1,.4}
\usepackage[normalem]{ulem}

\begin{document}
\preprint{AIP/123-QED}
\title{Stable amplitude chimera states in a network of locally coupled Stuart-Landau oscillators}
\author{K.  Premalatha}
\affiliation{Centre for Nonlinear Dynamics, School of Physics, Bharathidasan University, Tiruchirappalli - 620 024, Tamil Nadu, India.}
\author{V. K. Chandrasekar}
\affiliation{Centre for Nonlinear Science \& Engineering, School of Electrical \& Electronics Engineering, SASTRA University, Thanjavur -613 401,Tamilnadu, India.}
\author{M. Senthilvelan}
\affiliation{Centre for Nonlinear Dynamics, School of Physics, Bharathidasan University, Tiruchirappalli - 620 024, Tamil Nadu, India.}
\author{M. Lakshmanan}
\affiliation{Centre for Nonlinear Dynamics, School of Physics, Bharathidasan University, Tiruchirappalli - 620 024, Tamil Nadu, India.}
\date{\today}
\begin{abstract}
We investigate the occurrence of collective dynamical states such as transient amplitude chimera, stable amplitude chimera and imperfect breathing chimera states in a \textit{locally coupled} network of Stuart-Landau oscillators.  In an imperfect breathing chimera state, the synchronized group of oscillators exhibits oscillations with large amplitudes while the desynchronized group of oscillators oscillates with small amplitudes and this behavior of coexistence of synchronized and desynchronized oscillations fluctuates with time.  Then we analyze the stability of the amplitude chimera states under various circumstances, including variations in system parameters and coupling strength, and perturbations in the initial states of the oscillators.  For an increase in the value of the system parameter, namely the nonisochronicity parameter, the transient chimera state becomes a stable chimera state for a sufficiently large value of coupling strength.  In addition, we also analyze the stability of these states by perturbing the initial states of the oscillators.  We find that while a small perturbation allows one to perturb a large number of oscillators resulting in a stable amplitude chimera state, a large perturbation allows one to perturb a small number of oscillators to get a stable amplitude chimera state.  We also find the stability of the transient and stable amplitude chimera states as well as traveling wave states for appropriate number of oscillators using Floquet theory.  In addition, we also find the stability of the incoherent oscillation death states.   
\end{abstract}
\maketitle

\begin{quotation}
 Chimera states are complex spatio-temporal patterns where a network of identical coupled oscillators gets split into two coexisting regions of coherent and incoherent oscillations \cite{1,2}.  Initially it was assumed that nonlocal coupling is the necessary condition for the existence of chimera states in phase oscillators.  However, more recent studies reveal that systems with globally \cite{4, 4a, 4b, 4c} and locally coupled oscillators \cite{15, 16, 15b, 15a} are also capable of showing such phenomenon.  Recently, a new type of chimera state, namely amplitude chimera state was reported in a network of nonlocally coupled Stuart-Landau oscillators where the coherence and incoherence occur with respect to the amplitude alone while the phases of the oscillators are completely correlated \cite{20}.  This is in contrast with the amplitude-mediated chimera state where the chimera patterns exist with respect to amplitude and phase.  Such a state was reported in globally coupled Ginzburg-Landau oscillators \cite{4}.  Then the question arises whether simple local coupling can lead to amplitude chimera and other states.  To address this question, we here investigate the emergence of different dynamical states including transient amplitude chimera, stable amplitude chimera and imperfectly breathing chimera states in a locally coupled network of Stuart-Landau oscillators.  In addition, we analyze the stability of the amplitude chimera states under various circumstances, including variations in system parameters and coupling strength, and perturbations in the initial states of the oscillators.
\end{quotation}
\section{Introduction} 
\par The phenomenon of a hybrid type of spatiotemporal pattern, namely the coexistence of coherent and incoherent behaviours, has attracted much interest in the past decade.  This hybrid state was first observed in nonlocally coupled complex Ginzburg-Landau equation by Kuramoto and Battogtokh \cite{1}.  Later, it was named as chimera by Abrams and Strogatz \cite{2}.  Subsequently, it has been observed in coupled chaotic oscillators \cite{cha2}, time-discrete maps \cite{cha1}, neuronal networks \cite{5}, planar oscillators \cite{6}, networks with more than one population \cite{8, 9, 10,10a} and so on.  Experimental evidence for chimera states have also been found in chemical oscillators \cite{11}, opto-electronic \cite{12}, electro-chemical \cite{13}, mechanical \cite{14}, and electronic systems \cite{el}.  Chimera states have strong resemblance with many natural phenomena including epileptic seizure \cite{ep}, heart fibrillation \cite{h}, uni-hemispheric sleep \cite{u}, social systems \cite{s}, biological systems \cite{e}, SQUID meta materials \cite{sq}, etc.  
\par Initial studies showed that nonlocal coupling is essential for the existence of chimera states \cite{3}.  However recent studies reveal that a system of globally coupled oscillators can also have the capability to display such a phenomenon \cite{4, 41, 4a, 4b, 4c}.  Very recent works show that the restriction to observe the chimera state can be further relaxed to local coupling.  For instance, Laing has observed the chimera state in locally coupled reaction-diffusion equations in one dimension \cite{15}.  Later Li and Dierckx  \cite{16} have observed the existence of spiral wave chimeras in two dimensional locally coupled reaction-diffusion equations and Clerc {\it et al}.  \cite{15b} have shown that an ensemble of oscillators close to a homoclinic bifurcation can also exhibit chimera states.  Subsequently, in Ref. \cite{15a} Bera, Ghosh and Lakshmanan studied the existence of chimera states in local delay coupled oscillators.  They have also found that nonlinearity present in the local coupling can play an important role in the emergence of chimera states.  The discussed studies on chimera state dealt with local interaction involving highly nonlinear forms.  The question then arises whether simple local couplings, like linear ones, can lead to chimera states.  In the present study, we indeed demonstrate the existence of chimera states and study their stability under linear local coupling (that is the associated coupling term is a linear function of complex variable $z_j$, see Eq. (\ref{local}) below).

\par In this context, we also wish to note that the concept of amplitude chimera (AC) state has been studied by Zakharova {\it et al.} \cite{20} in a system of nonlocally coupled oscillators, where the chimera states occur with respect to the amplitudes of the oscillators while all the oscillators in the network are oscillating with the same frequency and correlated phase. Also the synchronized oscillators are oscillating periodically with the origin of the state space as a center of rotation while incoherent oscillators are oscillating periodically with a shifted center of rotation from the origin.  These authors have observed the amplitude chimera state in a network of nonlocal coupling with symmetry breaking coupling which is the crucial condition for the existence of such states.  This is in contrast with the amplitude-mediated chimera state observed in global coupling, where the chimera behavior is observed with respect to both amplitude and phase.  Later, amplitude chimera state was also reported in Ref. [27] with global coupling, and in Ref. [33] with nonlocal coupling where the coherent oscillators are oscillating with the same amplitude and incoherent oscillators are oscillating with different amplitudes but both the groups have the origin as a center of rotation.
\par Motivated by the above, in the present work we are interested to investigate the transient and stable amplitude chimera states as well as imperfect breathing chimera states in an array of  Stuart-Landau oscillators interacting via a \textit{linear local coupling}.  We identify a number of coupled dynamical states like transient and stable amplitude chimera states and imperfect breathing chimera state.  In an imperfect breathing chimera state, we observe that the synchronized group of oscillators are oscillating with large amplitudes, while the desynchronized group of oscillators are oscillating with small amplitudes and these behaviours repeat with time.  Then we analyze the stability of the amplitude chimera state with respect to various factors including the system parameters, coupling interaction and perturbation of initial states.  We find that the traveling wave solution is stable in the transient amplitude chimera state region.  For random initial conditions, we find that the completely synchronized solution is stable in the transient amplitude chimera state region.  In addition, we also find that the transient time of the amplitude chimera state increases for an increase of nonlocal coupling range from local coupling limit.  We also find the stability of the transient and stable amplitude chimera states as well as traveling wave states for appropriate number of oscillators using Floquet theory.  In addition, we also find the stability of the incoherent oscillation death states.  which are inhomogeneous steady states that alternatively occupy one of the two branches of the stable steady states.
\par The structure of the paper is as follows. In Sec. II, we introduce the model of locally coupled Stuart-Landau oscillators.  Then we discuss the different dynamical states under different initial conditions and investigate the dynamics of the amplitude chimera states and imperfect breathing chimera states.  In Sec. III, we present a detailed analysis of dynamical states under cluster initial conditions.  In Sec. IV, we analyze the stability of the amplitude chimera states and traveling wave states.  We summarize our findings in Sec. V.  

\section{Amplitude chimera states in locally coupled Stuart-Landau oscillators}
\subsection{Model}
To explore the nature of the collective states associated with the dynamics of a one-dimensional network of locally coupled oscillators, we consider the dynamical equations of the following system of Stuart-Landau oscillators,
\begin{eqnarray}
\dot{z_j}=(1+i\omega)z_{j}-(1- ic)|z_{j}|^2 z_{j}\nonumber\qquad \qquad\qquad\qquad\\
+\frac{\varepsilon}{2}(Re(z_{j-1})-2Re(z_{j})+Re(z_{j+1})),
\label{local}
\end{eqnarray}
where the dynamical variables $z_j=x_j+iy_j$ obey periodic boundary conditions $z_{j+N}=z_j$, $\omega$ is the natural frequency of the oscillators, $c$ is the nonisochronicity parameter and $j=1,2,3...N$, with $N$ being the total number of oscillators in the network.  In our simulation, we use the fourth order Runge-Kutta method with time step 0.01 to integrate the system (\ref{local}).  We generally leave out $10^8$ time units as transients in our analysis.  However, to identify certain transient chimera states, we also analyze certain states after leaving out smaller amount of transients which are explicitly mentioned in the text.  Here the coupling interaction is effected through the real part of the complex amplitude $z_j$ which breaks the rotational symmetry $z_j \rightarrow z_j^\prime=z_je^{i\theta}$ in the system.  To start with, we investigate the collective dynamical states associated with (1) for different sets of initial conditions and establish the existence of various types of chimera states.

\begin{figure*}[ht!]
\begin{center}
 \includegraphics[width=0.7\linewidth]{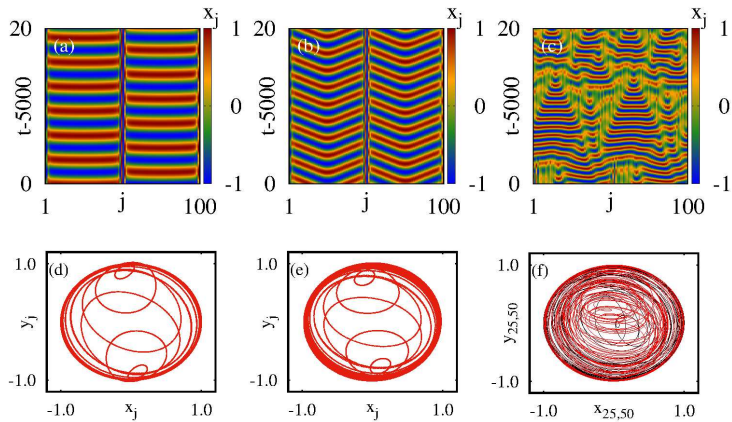}
\end{center}
\caption{(Color online) Spatio-temporal plots (a-c) and associated phase portraits (d-f), after leaving out $5\times10^3$ time units as transients: (a), (d) transient amplitude chimera (TAC) states for $c=0$, (b), (e)  stable amplitude chimera (SAC) states for $c=2$, (c), (f) imperfect breathing chimera (IBC) states for $c=5$.  Other parameter values: $\varepsilon=11$, $\omega=2$, $N=100$.}
\label{image1}
\end{figure*} 
\section{Study of the amplitude chimera states and imperfect breathing chimera states} 
\par In the above context, Zakharova et. al. \cite{20} have studied the existence of amplitude chimera states in a network of nonlocally coupled Stuart-Landau oscillators, and in Ref. [34] they have further analyzed these states with respect to initial conditions and transients.  In particular, they have reported that such chimeras exist for the above type of specific cluster initial condition where it was shown that a random shift from the initial states may significantly decrease the life time of the amplitude chimera states.  These authors have also found that for a random distribution without symmetries amplitude chimera states appear to be short transients towards in-phase synchronized region, while their lifetime may significantly increase for symmetric initial conditions. Further Tumash et al. \cite{flo} have noted that the existence of at least one positive real part of the Floquet exponents indicates an unstable manifold in phase space, which explains the nature of these states as long-living transients.
\par In the present study, we investigate the existence of transient amplitude chimera states and also analyze how such amplitude chimera states become stable with respect to an increase of nonisochronicity parameter.  To explore the spatio-temporal dynamics of system (\ref{local}) in some detail for the cluster initial conditions, we start by choosing the system parameter values as $\omega=2$ and $\varepsilon=11$.  Here the initial states of the oscillators are chosen as ($x_j,y_j$)=(+1,-1) for $j=1,2...N/2$ and ($x_j,y_j$)=(-1,+1) for $j=\frac{N}{2}+1,...N$.  Figs. \ref{image1} are plotted by leaving $5\times10^3$ time units as transients for three different values of the nonisochronicity parameter $c$.  By varying the strength of the coupling interaction, we can observe the existence of transient amplitude chimera state for the value $c=0$ which is shown in Fig. \ref{image1}(a).  Here the chimera state is transient and finally the system attains the traveling wave state (actually the transient nature of the amplitude chimera state is discussed in the following).  Amplitude chimera state represents the coexistence of two different domains: one is oscillating with spatially coherent amplitude while the other domain exhibits oscillations with spatially incoherent amplitudes, while the average phase velocity of each oscillator in the system remains the same.  The phase portrait of the oscillators in the amplitude chimera state is shown in Fig. \ref{image1}(d) which clearly illustrates that the synchronized oscillators are oscillating periodically with the origin of the state space as the center of rotation while the incoherent oscillators are oscillating periodically with different amplitudes and with a shifted center of rotation from the origin of the state space.  Increasing the value of nonisochronicity parameter to $c=2$ also leads to the existence of amplitude chimera state.  However, here the coherent domains are not completely synchronized where we can observe spatially two counter-moving domains starting from the middle of the coherent domain is illustrated in Figs. \ref{image1}(b) and (e).  This is different from the former case where the coherent domains are completely synchronized.  
\begin{figure}[ht!]
\begin{center}
 \includegraphics[width=1.0\linewidth]{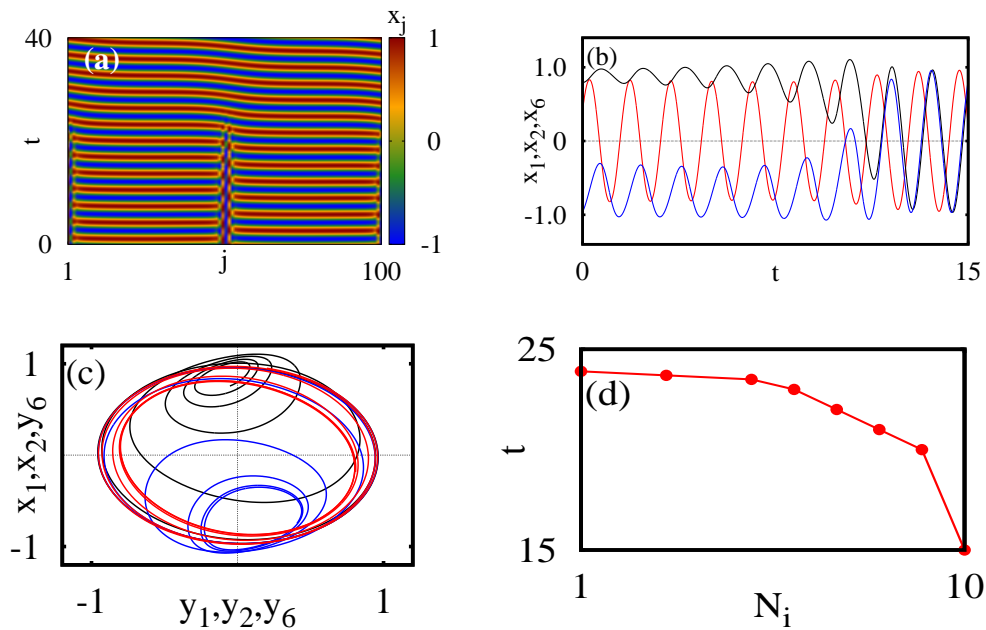}
\end{center}
\caption{(Color online) (a) Spatio-temporal plot for transient amplitude chimera state after leaving out $7660$ time units as transients, (b) time series of two chosen incoherent oscillators $(x_1, x_2)$ and one coherent oscillator $(x_6)$ (c) phase portraits of the corresponding oscillators $(z_1, z_2, z_6)$ and (d) the number of incoherent oscillators with respect to time.  Other parameter values: $\varepsilon=11$, $\omega=2$, $c=0$, $N=100$.}
\label{break}
\end{figure}
\begin{figure*}[ht!]
\begin{center}
 \includegraphics[width=0.7\linewidth]{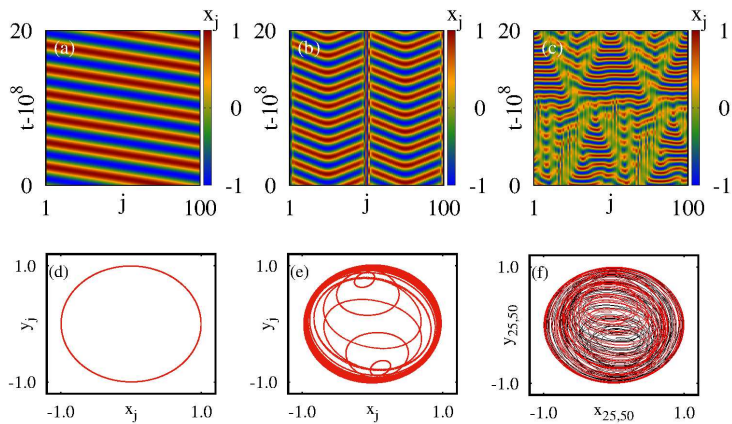}
\end{center}
\caption{(Color online) Spatio-temporal plots (a-c) and associated phase portraits (d-f), after leaving out $10^8$ time units as transients: (a), (d) traveling wave (TW) states for $c=0$ , (b), (e)  stable amplitude chimera (SAC) states for $c=2$, (c), (f) imperfect breathing chimera (IBC) states for $c=5$.  Other parameter values: $\varepsilon=11$, $\omega=2$, $N=100$.  (Note that the plots correspond to the same system considered in Fig. \ref{image1}, now evolved for a much longer time.)}
\label{image2}
\end{figure*}  
\par Interestingly, on increasing the value of $c$ further, the amplitude chimera state disappears and the large and small amplitude oscillations coalesce and result in an imperfect breathing chimera state as shown in Fig. \ref{image1}(c).  This state represents the fact that the coexistence of synchronized oscillations with large amplitude of oscillations and desynchronized oscillations with small amplitudes of oscillations persist for certain time, after which synchronized group becomes desynchronized and vice versa.  This tendency repeats with time but irregularly.  It is also to be noted that the oscillators belonging to incoherent group switch to coherent group, but not necessarily the same group of coherent/incoherent oscillators become incoherent/coherent.  Hence this state is designated as imperfect breathing chimera state.  This is different from the imperfect traveling chimera state where the incoherent traveling domain spreads into the coherent domain \cite{imperfect}.  For illustrative purpose, the phase portraits of two chosen oscillators $z_{25}$ (red/grey curve) and $z_{50}$ (black curve) are shown in Fig. \ref{image1}(f) which clearly show the existence of large amplitude of oscillations along with small amplitude oscillations.  In addition, we also analyze that how the transition occurs from amplitude chimera state to traveling wave state which is illustrated in Figs. \ref{break}(a-c).  The number of incoherent oscillators $N_i$ associated with the amplitude chimera state do not immediately combine into a traveling state which decreases with respect to time as shown in Fig. \ref{break}(d). 

\par
\begin{figure}[ht!]
\begin{center}
 \includegraphics[width=1.0\linewidth]{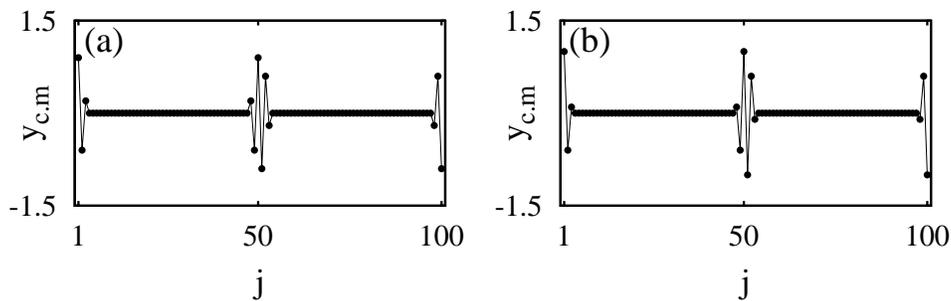}
\end{center}
\caption{Center of mass $y_{c.m}$ averaged over one period of each oscillator for the variable $y_j$: (a) transient amplitude chimera states for $c=0$ and (b) stable amplitude chimera states for $c=2$.  Other parameter values: $\varepsilon=11$, $\omega=2$, $N=100$.}
\label{center}
\end{figure}

 To confirm whether the dynamical states discussed in Figs. \ref{image1} are transient or stable, we plotted the space-time plot in Figs. \ref{image2} by leaving $10^8$ time units as transients for three different values of the nonisochronicity parameter $c$ (note that the values of $c$ are the same as in Figs. \ref{image1}) with $\omega=2$ and $\varepsilon=11$.  For the value of $c=0$, the system attains traveling wave state, after leaving out a long transient time ($10^8$ time steps) as in Fig. \ref{image2}(a) and the phase portraits of the oscillators are illustrated in Fig \ref{image2}(d). By increasing the value of the nonisochronicity parameter to $c=2$, we can observe the existence of stable amplitude chimera state which indicates that the AC state in Fig. \ref{image1}(b) is not transient.  Fig. \ref{image2}(b) shows the existence of stable amplitude chimera state, after leaving out a long transient time ($10^8$ time steps).   This is also confirmed by finding the center of mass $y_{c.m}=\int_0^{T} y_i(t)dt/T$, where $T=2\pi/\omega$ is the oscillation period for the $j$th oscillator for each of these cases. They are plotted corresponding to the transient and stable amplitude chimera states in Figs. \ref{center}(a) and (b), respectively.  From these figures we can clearly note that the oscillators in the coherent population are characterized by $y_{c.m}=0$, that is zero shift of center of mass from the origin, while the oscillators belonging to the incoherent group exhibit shifts in the position of the center of mass from the origin (here origin represents the origin of the state space $z$).  It can also be noted that in the case of transient amplitude chimera state, the oscillators consist of two completely synchronized domains and they are separated by an incoherent domain.  On the other hand, in the case of stable amplitude chimera states, the two coherently oscillating domains are separated by the incoherent domain.  Here, we can observe the spatially two counter-moving domains starting from the middle of the coherent domain.  If we look at the nature of dynamical states by leaving out long transient time of the order of $10^8$ units, we can find the traveling wave state in place of transient amplitude chimera state for $c=0$ as shown in Fig. \ref{image2}(a) and the phase portraits of the oscillators are shown in Fig. \ref{image2}(d).  Thus for the value of $c=2$, we can find the existence of amplitude chimera state even after leaving out a long transient time of the order of $10^8$ units and hence this state is designated as stable amplitude chimera state (Figs. \ref{image2}(b) and (e)).  Further increasing the value of the c to $c=5.0$, we can observe the existence of imperfectly breathing chimera states as in Fig. \ref{image2}(c) and the phase portraits of two randomly chosen oscillators $z_{25}$ and $z_{50}$ are shown in Figs. \ref{image2}(f).
 \begin{figure}[ht!]
\begin{center}
 \includegraphics[width=0.8\linewidth]{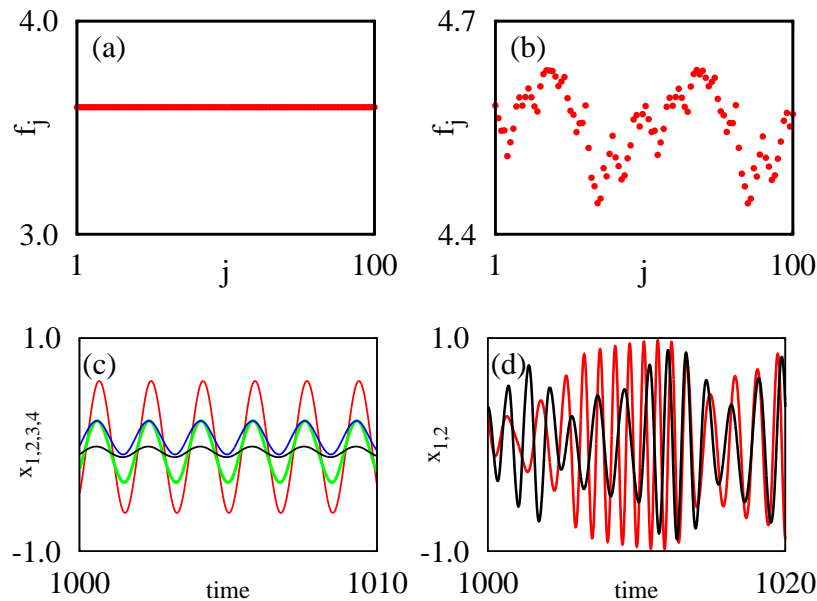}
\end{center}
\caption{(Color online) Frequency profile of the oscillators: (a) stable amplitude chimera states for $c=2$ (after leaving out $10^8$ time units as transients), (b) imperfect breathing chimera states for $c=5$ (after leaving out $10^8$ time units as transients).  Other parameter values: $\varepsilon=11$, $\omega=2$, $N=100$.  (c) Time evolution of the representative oscillators ($x_1,~x_2,~x_3,~x_4$) in stable amplitude chimera states, (d) time evolution of the representative oscillators ($x_1,~x_2$) in imperfect breathing chimera states.}   
\label{freq}
\end{figure}
    Another point to be noted here is that for small values of nonisochronicity parameter the amplitude chimera states are transient in nature. Then these states get stabilized for sufficiently increased values of $c$.  In the case of transient and stable amplitude chimera states the frequency of all the oscillators are the same while deviations occur with respect to their amplitudes.  Increase of nonisochronicity to further larger values of $c$ leads to deviations in both the amplitudes and frequencies which results in an imperfect breathing chimera state.  For illustrative purpose the average frequency profile of the oscillators in the stable amplitude chimera states are shown in Fig. \ref{freq}(a) for $c=2$.  The average frequency profile of the oscillators is calculated from the expression $f_j=2\pi\gamma_j/\Delta T$, where $j=2,3,...N$, and $\gamma_j$'s are the number of maxima in the time series $x_j$ of the $j^{th}$ oscillator during a sufficiently long time interval $\Delta T$ (here we consider the time interval as $\Delta T=5\times 10^8$ time units).  The time evolution of some of the representative oscillators in stable amplitude chimera states is plotted in Fig. \ref{freq}(c) which clearly illustrates the deviations in amplitudes even though the frequency of the oscillators are the same.  On the other hand, by increasing the value of nonisochronicity parameter to $c=5.0$, we can observe the deviations in the average frequency of the oscillators for an imperfect breathing chimera state which is illustrated in Fig. \ref{freq}(b).  The time evolution of the representative oscillators ($x_1,~x_2$) is shown in Fig. \ref{freq}(d).  We also observe from Figs. \ref{image1}(a, b) and \ref{image2}(a, b) that the oscillators in the amplitude chimera state display periodic oscillations, whereas individual oscillators in the imperfect breathing chimera state display chaotic behaviour.  The later behaviour of the oscillators is confirmed by using 0-1 test for one of the randomly chosen representative oscillator $z_{25}$ in this state.
\begin{figure}[ht!]
\begin{center}
 \includegraphics[width=1.0\linewidth]{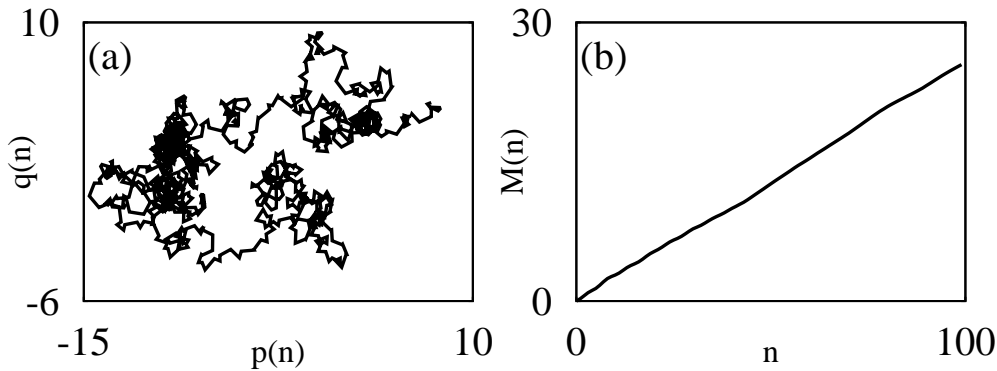}
\end{center}
\caption{(a) Dynamics of the translation variable $p(n)$ and $q(n)$ and (b) behaviour of mean square displacement $M(n)$ corresponding to the time series data of the randomly chosen representative oscillator $z_{25}$ (with 50000 data points) corresponding to imperfect breathing chimera states shown in Fig. \ref{image2}(f).}
\label{zero}
\end{figure} 
\subsection{0-1 test: Analysis of the temporal behaviour of the oscillators in an imperfect breathing chimera state}
\par The 0-1 test is used to classify the periodic/quasi-periodic/chaotic behaviours of the attractors.  In this test, the state of the oscillator is plotted in a space of translation variables which is given by 
\begin{eqnarray}
p(n)= \sum_{k=1}^{n}x(k) \cos kl, \quad q(n)=\sum_{k=1}^{n}x(k) \sin kl,
\end{eqnarray}
 where $x(k)$ is the times series data, $l$ is a fixed parameter chosen between $0$ and $2\pi$ (and in the present calculation it is fixed as 0.7, see Ref. [37]), and we have chosen $n=50000$.  The plot for $p(n)$ vs $q(n)$ is shown in Fig. \ref{zero}(a) which indicates the irregular or chaotic dynamics of the oscillators.  The diffusive or non-diffusive behaviour of the translation variables $p(n)$ and $q(n)$ can be analyzed by finding the mean square displacement $M(n)$ from the expression $\displaystyle M(n)=\lim_{N\to\infty}\frac{1}{N}\sum_{k=1}^{N}[p(k+n)-p(k)]^2+[q(k+n)-q(k)]^2.$  From Fig. \ref{zero}(b), we can observe that $M(n)$ is not bounded (that is linear with time).  Hence this confirms the temporal dynamics of the oscillators in the imperfect breathing chimera state corresponds to chaotic motion. 
\begin{figure}[ht!]
\begin{center}
 \includegraphics[width=1.0\linewidth]{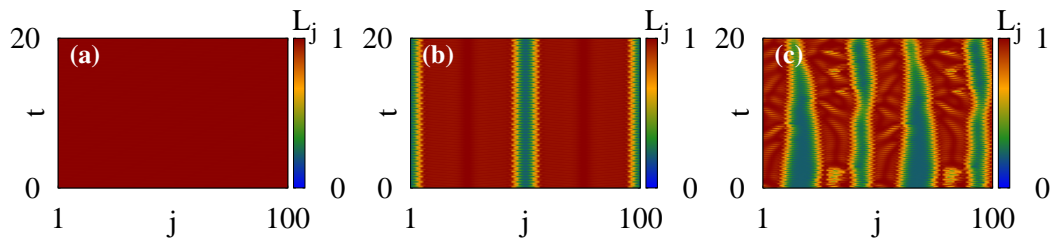}
\end{center}
\caption{(Color online) Local order parameter with node index $j$.  Red (gray) color indicates coherent and blue (dark gray) color represents incoherent domains.  (a) traveling wave states for $c=0$, (b) stable chimera states for $c=2$ and (c) imperfect breathing chimera states for $c=5$.  Other parameter values: $\varepsilon=11$, $\omega=2$, $N=100$.}
\label{order}
\end{figure}
\par To confirm that the pattern shown in Fig. \ref{image2} is indeed a chimera, we make use of the notion of the strength of incoherence $S$.  We find that the values of $S$ take the intermediate values between $0$ and $1$ for amplitude chimera states.  The details about the strength of incoherence are discussed briefly in appendix A.  In addition, we have also used local order parameter which represents the local ordering of the oscillators and thus the degree of (in)coherency.  This can be defined as \cite{local1,local2} $\displaystyle L_j=|\frac{1}{2\delta}\sum_{|j-k|\le\delta}e^{i\phi_k}|, \quad j=1,2,3,...N,$
where $\delta$ defines the nearest neighbors on both sides of the $j^{th}$ oscillator, $\phi_k=arctan(y_j/x_j)$.  The local order parameter of the $j^{th}$ oscillator, $L_j\approx 1$, indicates
that the $j^{th}$ oscillator belongs to the coherent part of the chimera state, that is, $L_j = 1$ means maximum ordering or coherency. On the other hand, $L_j\approx 0$ represents that the $j^{th}$ oscillator belongs to the incoherent neighboring nodes.  Here we take $\delta=6$ and compute the local order parameter $L_j$ of each oscillator for a long time interval which are shown in Fig. \ref{order}.  From Fig. \ref{order}(a), we can observe that $L_j = 1$ which implies the maximum ordering of the oscillators corresponding to traveling wave state.  Fig. \ref{order}(b) confirms the existence of chimera state where we can observe the coherent ($L_j = 1$) and incoherent domains ($L_j = 0$).  Local order parameter in Fig. \ref{order}(c) shows that the incoherent (and the coherent) domain is not static in time which means that coexistence of synchronized oscillations ($L_j = 1$) and desynchronized oscillations ($L_j = 1$) for certain time, after which synchronized group becomes desynchronized and vice versa.  This confirms the existence of imperfect breathing chimera state.  
\begin{figure}[ht!]
\begin{center}
 \includegraphics[width=0.8\linewidth]{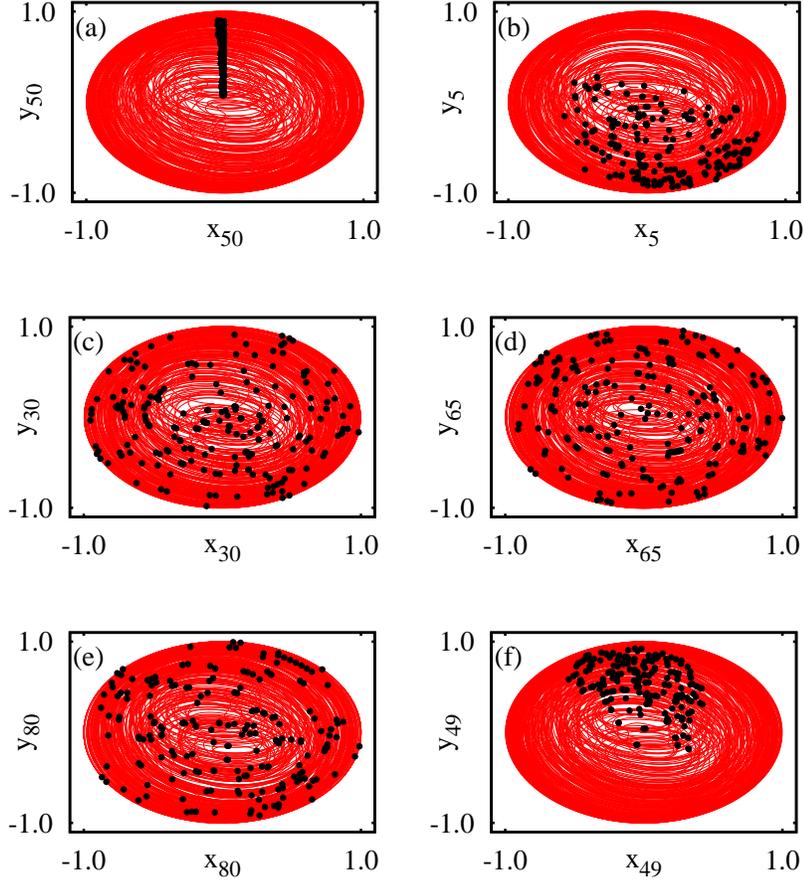}
\end{center}
\caption{(Color online) Localized sets obtained for the oscillators in the imperfect breathing chimera state: (a) represents the phase portrait of the reference oscillator with segment $\Lambda_{50}$ (black/grey dots) and (b-f) are the phase portraits of the randomly chosen oscillators with obtained data sets (black /grey dots).  Other parameter values are the same as for imperfect breathing chimera state in Fig. \ref{image2}(f).}
\label{locali}
\end{figure} 
\begin{figure}[ht!]
\begin{center}
 \includegraphics[width=1.0\linewidth]{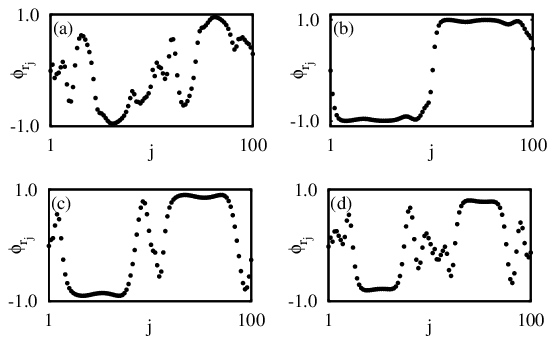}
\end{center}
\caption{(a)-(d) Relative phase $\phi_{r_{j}}$ of the oscillators for different intervals of time corresponding to the imperfect breathing chimera state shown in Fig. \ref{image2}(f).  Other parameter values are the same as in Fig. \ref{image2}(f).}
\label{relative}
\end{figure}
\subsection{Confirmation of imperfect breathing chimera state through localized set approach}
 \par In the imperfect breathing chimera state, some oscillations exist around the origin while some oscillations are not.  Also there exists a variation in the mean phase velocity of the oscillators.  Hence it is not meaningful to analyze the center of mass of each oscillator.  To analyze temporally the phase-locking behaviour among the coherent and incoherent oscillators, we make use of the localized set approach \cite{local}.  To illustrate this concept, we choose $z_{50}$ as a reference oscillator.  We construct the data set for randomly chosen 5 oscillators ($z_5, z_{30}, z_{49}, z_{65}, z_{80}$).  It is constructed by observing the data of these oscillators for every time the trajectory of the reference oscillator $z_{50}$ crosses the segment $\Lambda_{50}(x_{50},y_{50}$) (which is defined in the present case as $\Lambda_{50}=(x_{50} > 0.01, y_{50}\approx 0.8$).  Fig. \ref{locali}(a) shows the phase portrait of the reference oscillator with a specific segment $\Lambda_{50}$ (black/grey dots).  Figs. \ref{locali}(b-f) are the phase portraits of the randomly chosen oscillators and their corresponding observed data sets are given by black/grey dots along with their phase portraits.  It is clear from Figs. \ref{locali}(b) and (f) that some points in the data set are localized while some points are scattered over the trajectory which indicates that these oscillators are in-phase synchronized for certain time and for another time period they are not in-phase synchronized with the reference oscillator $z_{50}$.  On the other hand in Fig. \ref{locali}(c), (d) and (e), the data sets are spread over the trajectory which indicates that these oscillators are not in-phase synchronization with $z_{50}$.  If we look at the data set corresponding to the oscillator $z_{30}$, the data are spread over the phase trajectory,  eventhough the oscillator $z_{30}$ is part of the coherent group.  From this we conclude that due to the breathing nature of the chimera state, part of the group of oscillators show both phase locking (where the data sets are localized) and non-phase locking behavior (where the data sets are spread over the trajectory).  Also the remaining part of the oscillators shows complete non-phase locking behavior (where the data sets are completely spread over the trajectory).    
\par To get a better understanding of the phase dynamics of the oscillators in an imperfect breathing chimera state, we plotted the relative phase $\phi_{r_{j}}$ of the oscillators with respect to one of the representative oscillators as shown in Figs. \ref{relative}(a-d).  To find the relative phase $\phi_{r_{j}}$, we fix the segment $\Lambda_j$ $(x_j,y_j)$ in the phase space.  By considering the first oscillator as a reference oscillator, we note the position of all the other oscillators in the system, whenever the trajectory of the reference oscillator crosses the segment $\Lambda_j$.  Fig. \ref{relative} is plotted for $\phi_{r_{j}}$ with different intervals of time.  From Fig. \ref{relative}(a), we find the coexistence of spatially incoherent and coherent distributions of relative phases.  After certain time the system of oscillators gets split into two coherent domains with opposite phases as shown in Fig. \ref{relative}(b).  While looking at the distribution of the oscillators after some time, we find the coexistence of coherent and incoherent distributions of relative phases as shown in Fig. \ref{relative}(c) and (d).  Similarly, we can observe the same tendency alternatively for different time intervals but irregularly with time.  It can be seen from Figs. \ref{relative}(a-d), that the oscillators belonging to the incoherent group switch to coherent group but not necessarily the same group of coherent/incoherent oscillators become incoherent/coherent.  This confirms the breathing behaviour of chimera states which was shown in Fig. \ref{image2}(f).  We can also observe that the relative phase of the oscillator $z_{30}$ is a part of the coherent group, even if the data sets observed from the localized set approach for such oscillator are randomly distributed over the trajectory (indicating a non-phase locking behaviour of the oscillator $z_{30}$ with the reference oscillator as shown in Fig. \ref{locali}(c)).  It can also be noted that the coherent and incoherent domains of the relative phases of the oscillators are shifting spatially with time.
\section{Collective dynamics under two different initial conditions: Existence of amplitude chimera states and imperfect breathing chimera states}
In this subsection, we discuss the global picture of the system (\ref{local}) for two specific initial conditions: (i) cluster initial conditions, (ii) random initial conditions.  In particular, we deduce the two parameter phase diagrams in the ($\varepsilon, ~c$) plane for a fixed value $\omega=2$, and $N=100$, for both sets of initial conditions.  The dynamics of the system for larger system size $N$ is discussed latter in section-V. 
\begin{figure*}[ht!]
\begin{center}
 \includegraphics[width=1.0\linewidth]{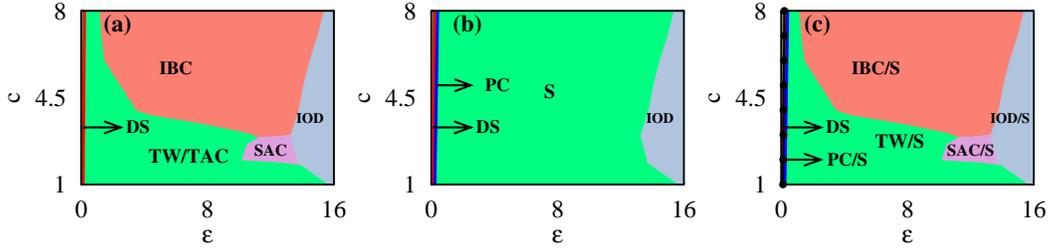}
\end{center}
\caption{(Color online) Two parameter phase diagrams in the ($\varepsilon, c$) plane for 100 oscillators: (a) Dynamical states for cluster initial conditions, (b) dynamical states for random initial conditions and (c) dynamical states with multistability regions (which is the overlay of the dynamical states observed for cluster initial condition (Fig. (a)) as well as for random initial conditions (Fig. (b))).  In Fig. (a) DS represents the desynchronized state, TW represents the traveling wave, TAC is the transient amplitude chimera state, SAC is the stable amplitude chimera state, IBC represents the imperfect breathing chimera states, IOD is the incoherent oscillation death states.  In Fig. (b), PC represents phase chimera state.  In Fig. (c) region-TW/S is the multistability region between the traveling wave and completely synchronized state. Region-PC/S is the multistability region between the PC (phase chimera) state and completely synchronized state, regions IBC/S and SAC/S are multistability regions between the completely synchronized state and the imperfect breathing chimera state and stable AC state, respectively. Region-IOD/S is the multistability region between the IOD and completely synchronized state.  Dotted line between the desynchronized and other dynamical regions (in Fig. (c)) represents the stability curve estimated from the Lyapunov exponents of the variational equation (\ref{vari}).}
\label{two}
\end{figure*}
\subsubsection{Cluster initial conditions}
\par
\begin{figure}[ht!]
\begin{center}
 \includegraphics[width=1.0\linewidth]{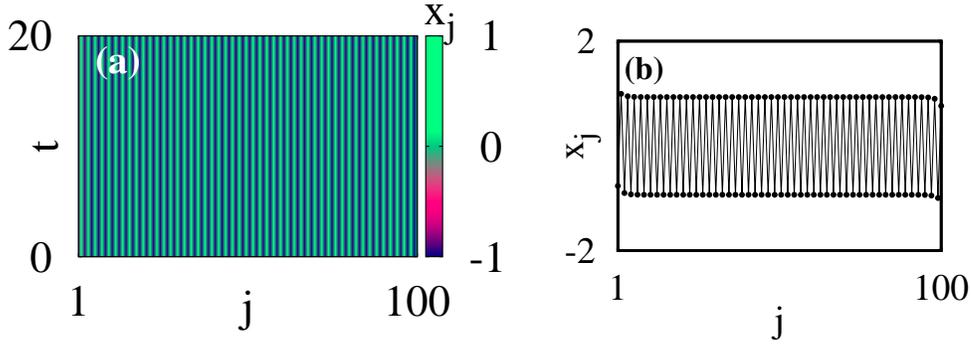}
\end{center}
\caption{(a) Spatio-temporal plot of the variables $x_j$ for incoherent oscillation death state and (b) the corresponding snapshot of the variables $x_j$ at $t=15$ with the parameter values $c=3$ and $\varepsilon=16$.}
\label{icod}
\end{figure}

 In order to know the different dynamical regimes of the system (\ref{local}) for cluster initial conditions, we plotted the two parameter phase diagram in the ($\varepsilon, ~c$) plane in Fig. \ref{two}(a).  Here the initial states of the oscillators are chosen as ($x_j,y_j$)=(+1,-1) for $j=1,2...N/2$ and ($x_j,y_j$)=(-1,+1) for $j=\frac{N}{2}+1,...N$.  To start with we find that the system of oscillators is in a state of phase desynchronization for small values of coupling strength.  Phase desynchronization here represents the state where the oscillators are oscillating with same amplitude and frequency while their phases are different.  Numerical boundary of regions of phase desynchronization (denoted by DS in Fig. \ref{two}(a)), synchronization and amplitude chimera states are identified with the help of strength of incoherence \cite{21} (for more details see appendix).  From Fig. \ref{two}(a), we find that the stable amplitude chimera state (SAC) exists only in a small region of the parametric space.  By amplitude chimera state we mean here the coexistence of two different domains: one is oscillating with spatially coherent amplitude while the other domain exhibits oscillations with spatially incoherent amplitudes, where the average phase velocity of each oscillator in the system remains the same.  In the region corresponding to transient amplitude chimera (TAC) state, traveling wave (TW) solutions are stable.  Here the amplitude chimera state emerges only in the transient time.  That is the life time of the amplitude chimera state is finite.  In the asymptotic limit, the amplitude chimera disappears while the traveling wave state exists in this region.  By increasing the nonisochronicity parameter, we find large regions of imperfect breathing chimera state (IBC).  The stable amplitude chimera (SAC) state region (observed for the cluster initial condition) is the bistable region for the reason that random initial conditions in this region leads to the existence of complete synchronized solution, as pointed out in the next subsection. For any choice of c, one finds that a large value of coupling interaction leads to incoherent oscillation death (IOD) due to symmetry breaking in the coupling.  Here the inhomogeneous steady states alternatively occupy one of the two branches of the stable steady state which results in the incoherent oscillation death state.  This is different from coherent oscillation death (where neighboring oscillators populate in the same branch (in either upper branch or lower branch) of the inhomogeneous steady state).  This is clearly shown in Fig. \ref{icod}(a) while a snapshot of the variables at $t=15$ is presented in Fig. \ref{icod}(b).
\subsubsection{Random initial conditions} 
\par The dynamical regions which exist in the system (\ref{local}) for random initial conditions are illustrated in Fig. \ref{two}(b), where we also compare with the states which one obtains for cluster initial conditions.  We choose random initial conditions between $-1$ to $+1$ for every choice of $\varepsilon$ and $c$ values. For this purpose, we scan the $(\varepsilon, c)$ plane with increments of 0.015 in $\varepsilon$ and 0.045 in $c$.  For sufficiently small values of coupling strength $\varepsilon$, the system of oscillators are oscillating incoherently.  By increasing the value of $\varepsilon$, the system attains incoherent oscillation death (IOD) through the phase chimera state and synchronized state for all values of nonisochronicity parameter.  
\subsubsection{Multistability states} 
\begin{figure}[ht!]
\begin{center}
 \includegraphics[width=1.0\linewidth]{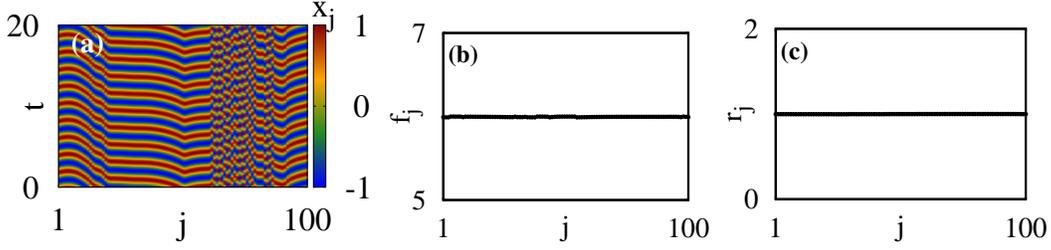}
\end{center}
\caption{(a) Spatio-temporal plot of the variables $x_j$ for phase chimera state, (b) the corresponding frequency of the oscillators and (c) the amplitude of the oscillators with the parameter values $c=3$ and $\varepsilon=0.1$.}
\label{phasechimera}
\end{figure}

Figure \ref{two}(c) illustrates the multistability regions which exist in the system (\ref{local}).  In the region TW/S, we can observe the existence of completely synchronized solution for randomly chosen initial conditions between -1 to +1 for the oscillators ($x_j,~y_j$).  On the other hand in this region, for cluster initial conditions we can observe the onset of amplitude chimera states which are transient while traveling wave solutions are stable.  In region PC/S, we can observe the existence of phase chimera state while the synchronized solution is stable for the choice of initial state of the oscillators near synchronized state.  Here the phase chimera state represents the coexistence of coherent and incoherent distributions only in the phases of the oscillators while their amplitudes and mean phase velocities remain the same which are clearly illustrated in Fig. \ref{phasechimera}.  In Ref. [33], the present authors reported amplitude chimera states in the case of nonlocal coupling for random initial conditions between -1 and +1.  In such a case, the coherent oscillators are oscillating with same amplitude while the incoherent oscillators are oscillating with different amplitudes but both the groups have the origin as the center of rotation which is different from the amplitude chimera state reported in the present study.  Also in this region, we can observe the transient amplitude chimera/traveling wave state for the cluster initial condition.  Similarly, in region IBC/S and SAC/S, imperfect breathing chimera states and stable amplitude chimera states are stable for cluster initial condition.  For randomly chosen initial conditions, we can find the completely synchronized states to be stable. Also in region IOD/S, incoherent oscillation death (IOD) states coexist with the synchronized state.  That is in this region we can observe the IOD state to be stable for cluster initial condition and synchronized state is stable if initial states of the oscillators are chosen away from the cluster initial condition.  Thus we conclude that random initial conditions support the synchronized solution and suppresses the amplitude chimera region. 

\subsubsection{Stability of the synchronized state using master stability function}
 We have next identified the boundary between the desynchronized state and synchronized state in both the Figs. \ref{two}(a) and \ref{two}(b) using the notion of master stability function (MSF) \cite{msf1,msf2,msf3}.  The stability of the synchronized solution ($x_j=x$, $y_j=y$, $\forall j$) is determined by the variational equations 
\begin{eqnarray}
\dot{\eta}_{1j}=(1-3x^2-2cx y -y^2)\eta_{1j}\nonumber\qquad \quad \quad \\-(\omega+cx^2+3cy^2+2x y )\eta_{2j}+\varepsilon\lambda_j\eta_{1j},\nonumber\\
\dot{\eta}_{2j}=(\omega+3cx^2-2x y +cy ^2)\eta_{1j}\nonumber\qquad \quad \quad\\
+(1+2cx y -x^2-3y^2)\eta_{2j},~~j=1,2,... N,
\label{vari}
\end{eqnarray}
where $x(t)$ and $y(t)$ are the solutions of the uncoupled system (1) corresponding to the synchronized motion, $\eta_{ij}$'s, $i=1,2,$ are the perturbations from the synchronized manifold and $\lambda_j$'s are the eigenvalues of the coupling matrix.  Then the eigenvalues are given by 
\begin{eqnarray}
\lambda_j=-1+\cos (\frac{2\pi}{N}j), ~~ j=0,1,2,... N-1.
\end{eqnarray}
 The eigenvalue $\lambda_0$ corresponds to the perturbation parallel to the synchronized manifold.  Other $N-1$ eigenvalues are associated with the transverse manifold.  The transverse eigenmodes should be damped out to get a stable synchronization manifold.  The Lyapunov exponents corresponding to the variational equation with $\lambda_j$ determine the stability of the synchronous state.  Suppose that the variational equation with $\lambda_j$ gives Lyapunov exponents $\zeta_1^{(j)} \geq \zeta_2^{(j)}$. Then the stability of the synchronous state requires $\zeta_1^{(j)}< 0$ for all $j$ ranging from 1 to $N-1$.  In Fig. \ref{two}(c), the dotted line is computed when all $\zeta_1^{(j)}$'s turn out to be negative.  For the region left of the dotted line the synchronous state is unstable and to the right of the dotted line the synchronous state is stable.   
\subsubsection{Stability of the amplitude chimera states and traveling wave states using Floquet theory and the incoherent oscillation death states}
\par
\begin{figure}[ht!]
\begin{center}
 \includegraphics[width=1.0\linewidth]{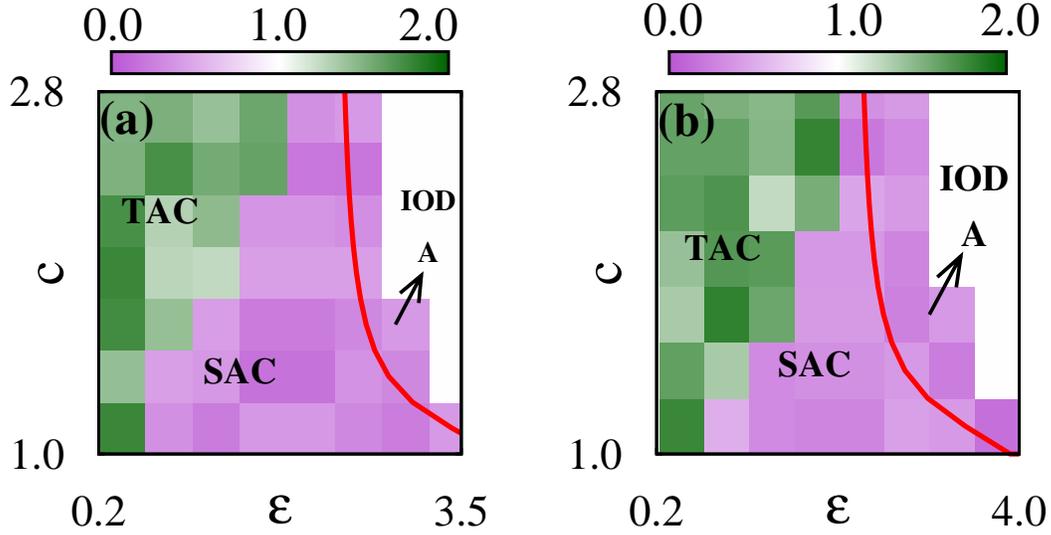}
\end{center}
\caption{(Color online) Largest value of Floquet multiplier $|\mu|_{max}$ (excluding the Goldstone mode) in the $(\varepsilon,c)$ plane: (a) for an array of oscillators $N=10$, (b) for number of oscillators with $N=14$.  TAC represents the transient amplitude chimera state and SAC represents the stable amplitude chimera state.  $|\mu|_{max}>1$ indicates the unstable nature of the periodic orbit in amplitude chimera state and  $|\mu|_{max}<1$ represents the stable nature of the periodic orbit.  IOD represents the incoherent oscillation death states.  Red/grey line is the stability curve for the IOD state which closely matches with the numerically obtained boundary for the initial condition near the IOD states.  Region-A represents the multistability region between the incoherent oscillation death state and stable amplitude chimera state.}
\label{floq}
\end{figure}

 Next, the stability of the amplitude chimera state is analyzed by studying the local stability of the periodic solution using Floquet theory \cite{flo}.  For this purpose, we linearize the system of equations (\ref{local}) by perturbing the solutions and we get 
{\small \begin{eqnarray}
\dot{\eta_j}=(1-3x_j^2-y_j^2+2cx_jy_j)\eta_j\qquad\qquad \qquad\qquad\qquad \qquad\qquad\qquad\\\nonumber
-(\omega+cx_j^2+3cy_j^2+2x_jy_j)\xi_j+\frac{\varepsilon}{2}(\eta_{j+1}-2\eta_j+\eta_{j-1})\qquad\qquad\qquad\\\nonumber
\dot{\xi_j}=(\omega+3cx_j^2-2x_jy_j+cy_j^2)\eta_j\qquad \qquad\qquad\qquad \qquad\qquad\qquad\quad\\\nonumber
+(1+2cx_jy_j-x_j^2-3y_j^2)\xi_j. \qquad\qquad\qquad\qquad\qquad\nonumber
\end{eqnarray}}
Integrating the above equation for one time period $T=\frac{2\pi}{\nu}$, where $\nu$ is the frequency of the periodic orbit, we can construct the monodromy matrix.  Then the eigenvalues of the monodromy matrix give rise to the Floquet multipliers $\mu_j$ \cite{flo}.  Here, we can observe the occurrence of amplitude chimera even for a minimal number of oscillators, say $N=10$ (essentially to simplify the analysis, which is then extended to a higher number of oscillators, see below).  If all the eigenvalues (Floquet multipliers) of the matrix $|\mu_j|$ are less than one (except for the Goldstone mode), $j=1,2,...10$, the corresponding periodic orbit is stable.  For the periodic orbits there always exists one Floquet multiplier $|\mu|=1$ which corresponds to the Goldstone mode.  If among the remaining multipliers, even one of the $|\mu_j|>1$ it signifies that the perturbation increases exponentially and that the periodic orbit is unstable.  Fig. \ref{floq}(a) shows the largest value of Floquet multiplier $|\mu|_{max}$ (excluding the Goldstone mode) in the $(\varepsilon,c)$ plane by leaving out $500$ time units as transients for $N=10$.  The value of $|\mu|_{max}$ is observed from the Floquet multiplier using the fundamental matrix.  In the figure $|\mu|_{max} >1$ corresponds to the unstable periodic orbit while $|\mu|_{max}<1$ represents the stable periodic orbit.
\begin{figure}[ht!]
\begin{center}
 \includegraphics[width=1.0\linewidth]{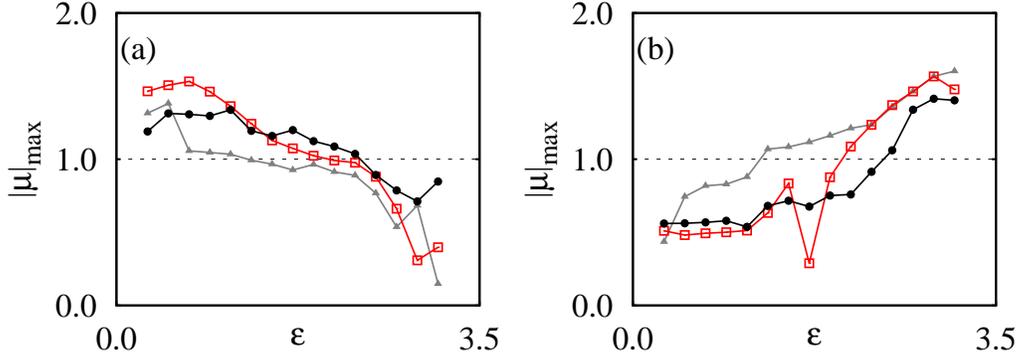}
\end{center}
\caption{(Color online) Maximum value of $|\mu|_{max}$ as a function of $\varepsilon$ for an array of oscillators with $N=10$: (a) for stability change of periodic orbit of amplitude chimera states, (b) for stability change of periodic orbit of traveling wave solution. Dark grey curve with ($\blacktriangle$) is for $c=1.5$, red/grey curve with $(\square)$ is for $c=2.0$ and black curve with $(\large{\bullet})$ is for $c=2.4$.   $|\mu|_{max}>1$ indicates the unstable nature of the periodic orbit and  $|\mu|_{max}<1$ represents the stable nature of the periodic orbit.  The critical value of coupling strength $\varepsilon_c$ (at which $|\mu|_{max}$ crosses the value 1) for both panels (a) and (b): $\varepsilon_c=1.3$ for $c=1.5$, $\varepsilon_c=2.1$ for $c=2.0$ and $\varepsilon_c=2.5$ for $c=2.4$.}
\label{real}
\end{figure} 
 We have also verified the stability of the amplitude chimera states by increasing number of oscillators in a network with $N=14$  and the corresponding maximum value of the Floquet multiplier $|\mu|_{max}$ in the ($\varepsilon, c$) plane is shown in Fig. \ref{floq}(b).  One thing to be noted here is that in the case of networks with smaller number of oscillators (that is $N=10,~14$), we can observe the existence of stable amplitude chimera state even for small values of coupling interaction compared to the network with a large $N$ (as in Fig. 1).  Hence there occurs quantitative changes in the dynamical regions, depending on the size of the system \cite{carol}.  Due to the presence of local coupling, even for small values of system size we can observe the same dynamical regions by increasing the strength of the coupling interaction. Such type of coupling scheme can be realized in neural systems \cite{15a}, lattice models \cite{shi}, etc.  In Fig. \ref{real}(a), we illustrate how the stability of the amplitude chimera state varies with an increase of coupling strength by plotting the maximum value of Floquet multiplier $|\mu|_{max}$ for three different values of nonisochronicity parameter with $N=10$.  Here the maximum value of $|\mu|_{max}>1$ for transient amplitude chimera state while $|\mu|_{max}<1$ for stable amplitude chimera state.  
\par We have also analyzed the stability of the periodic orbit associated with the traveling wave solution.  Traveling wave solution can be written in the form as $z_j(t)=\rho e^{i(\lambda t+2\pi jk/N)}$, where $\rho,\lambda, k \in\mathbb{R}$ are constant parameters.  Here $\rho$ is the amplitude,  $\lambda$ is the frequency and $k$ is the wavenumber.  Stability nature of the periodic orbit associated with the traveling wave solution is illustrated in Fig. \ref{real}(b) for $N=10$.  Maximum value of $|\mu|_{max}<1$ indicates the stability of the periodic orbit of the traveling wave solution  while $|\mu|_{max}>1$ indicates the unstable nature of the traveling wave solution.  In Figs. \ref{real}(a) and (b), dark grey curve with ($\blacktriangle$) is plotted for $c=1.5$, red/grey curve with $(\square)$ is plotted for $c=2.0$ and black curve with $(\Large{\bullet})$ is plotted for $c=2.4$.  We have checked the existence of stable amplitude chimera states for a network with N = 100 numerically by leaving long transient times as $10^8$ time units. 
\par  On the other hand, incoherent oscillation death (IOD) state represents the situation where the total population is split into two groups of inhomogeneous steady states.  In such steady states, neighboring oscillators alternatively occupy among the two distinct values of stable steady states.
The system has equilibrium points $(x_j,y_j)$=$(x_0,y_0$), $(x_{j+1},y_{j+1})$=$(-x_0,-y_0$) and $(x_{j-1},y_{j-1})$=$(-x_0,-y_0$) (as oscillators are distributed in IOD states).  Hence Eq. (\ref{local}) can be reduced as 
\begin{eqnarray}
x_0-\omega y_0-(x_0+cy_0)(x_0^2+y_0^2)-2\varepsilon x_0=0\\
\omega x_0+y_0-(y_0-cx_0)(x_0^2+y_0^2)=0.
\label{death}
\end{eqnarray}
The above equation has the explicit fixed point solution as 
\begin{eqnarray}
x_0= -\frac{\sqrt{\alpha-\beta-c^2\varepsilon(-2+3\varepsilon+\omega^2)}}{\sqrt{2}\varepsilon(1+c^2)},
\label{f1}
\end{eqnarray}
\begin{eqnarray}
y_0=\frac{(\gamma\varepsilon^2+\beta)x_0}{\sqrt{2}\varepsilon \gamma(c+\omega)},
\label{f2}
\end{eqnarray}  with {\small $\alpha=-2 \varepsilon^3 + \varepsilon^2 (1 - 3 c^2 - 4 c2 \omega)-\varepsilon(c+ \omega) (-2 c+ (-1 + c^2) \omega)$}, $\small{\beta=\sqrt{\varepsilon^2 \gamma^2 (\varepsilon^2 + c^2 (-1 + 2 \varepsilon) + 2 c (-1 +\varepsilon) \omega - \omega^2)}}$ and $\gamma=(-1 + c(c+ 2\omega) + 2 \varepsilon )$.
The stability curve observed for the incoherent oscillation death state using the above fixed point matches with the numerical boundary observed for the initial condition chosen near the incoherent oscillation death state which is illustrated with red/grey line in Fig. \ref{floq}(a) for $N=10$ and in Fig. \ref{floq}(b) for $N=14$.  We can find the multistability between the stable amplitude chimera state and incoherent oscillation death state in region-A.  In this region, we can observe the stable amplitude chimera state for the choice of cluster initial condition.  Here, we can also find that the synchronized solution coexists for the initial condition near to it.
\section{Stability of amplitude chimera state}
\par
\begin{figure}[ht!]
\begin{center}
 \includegraphics[width=1.0\linewidth]{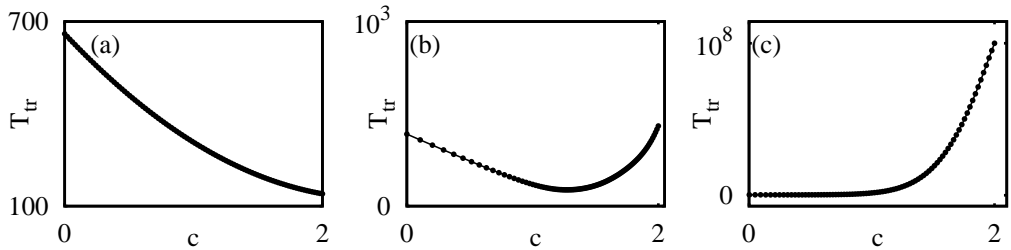}
\end{center}
\caption{Transient time $T_{tr}$ of amplitude chimera state versus nonisochronicity parameter $c$ for three different fixed values of coupling strength: (a) $\varepsilon=5$, (b) $\varepsilon=8$, (c) $\varepsilon=11$. }
\label{fit}
\end{figure}

 In the previous section, we have studied the existence of amplitude chimera state and noted that this state transits to imperfect breathing chimera state through stable amplitude chimera state while varying the nonisochronicity parameter.  There arises a question as to how the variation in the nonisochronicity parameter affects the stability of the amplitude chimera state.  To answer this question, we plotted the transient time of the dynamical state as a function of the nonisochronicity parameter ($c$) for three different fixed values of coupling strength in Fig. \ref{fit}.  From this figure we can observe that for small values of coupling strength ($\varepsilon=5$) the life-time of the amplitude chimera state decreases for an increase of $c$ as shown in Fig. \ref{fit}(a).  In this region, the amplitude chimera state is transient while traveling wave solution is stable.  On the other hand, on increasing the value of $\varepsilon$ to $\varepsilon=8$, the transient time $T_{tr}$ decreases with an increase of $c$, and then after a particular value of $c$, $T_{tr}$ again increases with an increase of $c$ as shown in Fig. \ref{fit}(b).  Here also the amplitude chimera state is transient but only the life time of this state increases.  On further increasing the value of $\varepsilon$ to $\varepsilon=11$, we find that the life-time of the chimera state gradually increases and then the transient state becomes stable (Fig. \ref{fit}(c)).  Thus we conclude that for small values of coupling strength the amplitude chimera state is transient and traveling wave solution is stable.  If the coupling strength is sufficiently large then the transient chimera state becomes a stable amplitude chimera state while increasing the parameter $c$. 
\begin{figure}[ht!]
\begin{center}
\includegraphics[width=0.8\linewidth]{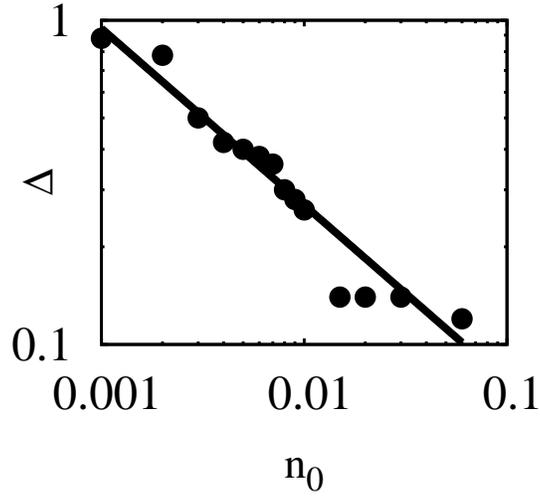}
\end{center}
\caption{Log-log plot for the value of perturbation ($\Delta$) and critical value $n_0$ corresponding to stable amplitude chimera state.  Other parameter values: $\varepsilon=11$ and $c=2$.}
\label{log}
\end{figure}
\begin{figure}[ht!]
\begin{center}
 \includegraphics[width=1.0\linewidth]{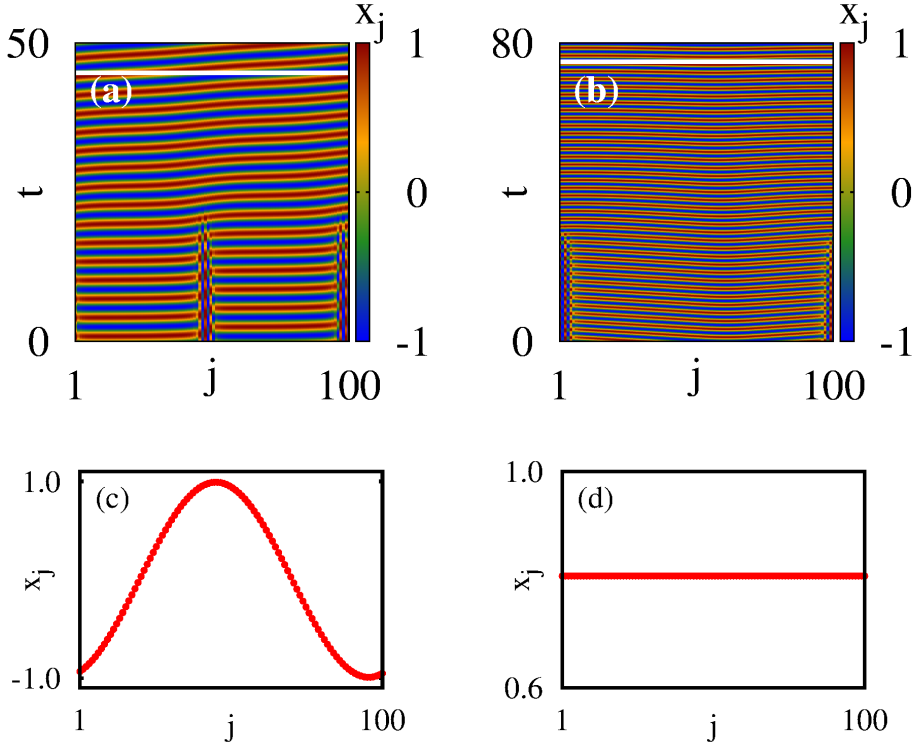}
\end{center}
\caption{(Color online) Spatio-temporal plot for the variables $x_j$: (a) traveling wave state for $c=0$ and $\varepsilon=11$ after leaving out the transient time $156 \times 10^3$ units for cluster initial conditions, (b) complete synchronization for $c=0$ and $\varepsilon=11$ while perturbing the initial states of the oscillators in transient amplitude chimera states after leaving out the transient time $220$ units.  Snapshots for the variables $x_j$: (c) for the traveling wave state at $t=45$ which is marked by the white line in Fig. \ref{stability}(a), (d) for the synchronized state at $t=77$ which is marked by the white line in Fig. \ref{stability}(b). }
\label{stability}
\end{figure}
\begin{figure}[ht!]
\begin{center}
 \includegraphics[width=1.0\linewidth]{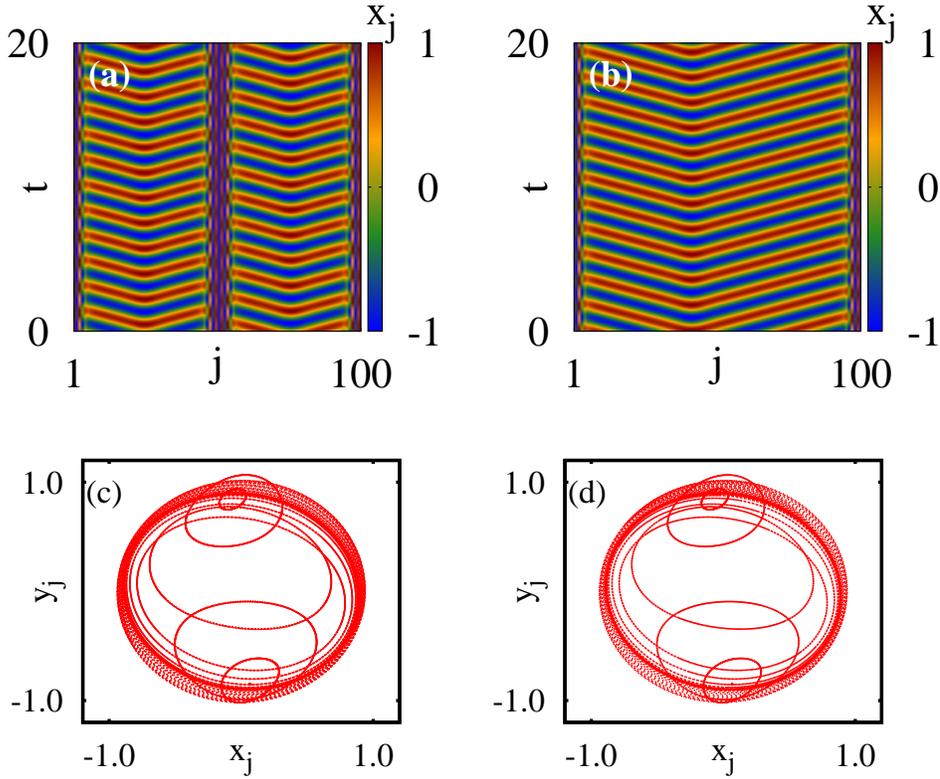}
\end{center}
\caption{(Color online) Spatio-temporal plot (a-b) and associated snapshots (c-d) for the variables $x_j$ in the amplitude chimera state after leaving out $10^8$ time units as transients: (a), (c) perturbation of small number of oscillators $N_{\epsilon}=5$, (b), (d) perturbation of large number of oscillators $N_{\epsilon}=85$.  Other parameter values: $\varepsilon=11$ and $c=2$.}
\label{stable}
\end{figure}

\begin{figure}[ht!]
\begin{center}
 \includegraphics[width=1.0\linewidth]{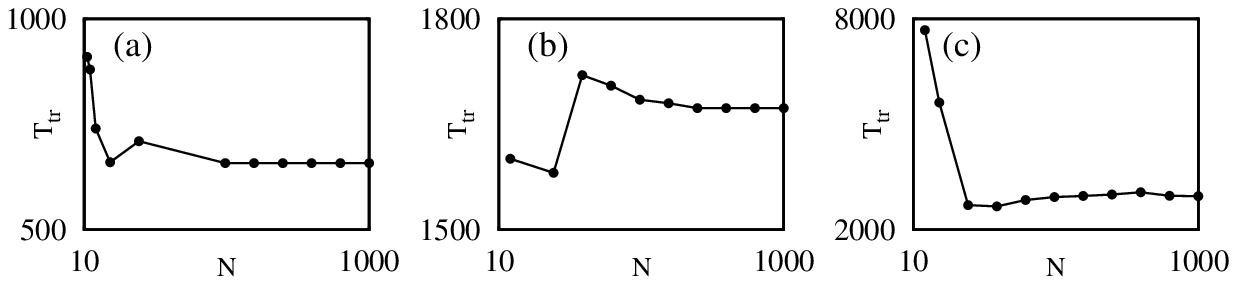}
\end{center}
\caption{Transient time of amplitude chimera state as a function of system size $N$: (a) for $\varepsilon=5$, (b) $\varepsilon=8$ and (c) $\varepsilon=11$ with $c=0$.}
\label{number}
\end{figure}  
\subsection{Stability with respect to perturbation of initial condition} 
In the previous section we chose the initial condition as ($x_j,y_j$)=(+1,-1) for $j=1,2,...,N/2$ and ($x_j,y_j$)=(-1,+1) for $j=\frac{N}{2}+1,...,N$. To analyze the stability, we start perturbing the initial state of the oscillators with ($x_j-j\Delta,y_j+j\Delta $) for $j=1,2,...,N/2$ and ($x_j+j\Delta,y_j-j\Delta$) for $j=\frac{N}{2}+1,...,N$.  Note that we start to perturb the oscillators in the ending edge for the first cluster ($j=1,2,...,N/2$) and perturb the oscillators in the starting edge for the second cluster ($j=\frac{N}{2}+1,...,N$). The chosen value of the number of oscillators from both the clusters are the same.  The value of $\Delta$ is an integral multiple of order $0.001$ (that is $\Delta=j\times 0.001$, $j=1,2,...,N$), that is the chosen number of oscillators are distributed with uniform difference $\Delta=0.001$.  The relation between the perturbation ($\Delta$) and critical value ($n_0=\frac{n}{N}$) is plotted in a log-log scale in Fig. \ref{log}.  Here $n$ represents the number of perturbed oscillators.  From this, we can observe that when the amount of perturbation is small it allows one to perturb more number of oscillators from the cluster initial condition.  If the perturbation is large, it allows one to perturb only a small number of oscillators.  Also the perturbation value ($\Delta$) follows the power law relation $\Delta=pn_0^q$ with $n_0$.  If the chosen number of oscillators is greater than $n_0$, ultimately the system of oscillators enters into a completely synchronized state.  In the transient amplitude chimera state region, traveling wave solutions are stable, and such a state is shown in Fig. \ref{stability}(a).   On the other hand, perturbation in the initial states of the oscillators leads to the completely synchronized solution as shown in Fig. \ref{stability}(b).  Figs. \ref{stability}(c) and \ref{stability}(d) show the snapshots of the variables $x_j$ for traveling wave state and synchronized state, respectively.  Fig. \ref{stable}(a) shows the stable amplitude chimera states by perturbing 5 oscillators in each cluster with $\Delta=0.001$.  From this figure, we can observe the incoherent domains in the edges as well as in the middle of the coherent domains and the phase portraits of the oscillators are shown in Fig. \ref{stable}(c).  If we increase the number of perturbed oscillators, we can observe the disappearance of incoherent domain in the middle of the coherent domain and such incoherent oscillators evolve with coherent group as shown in Fig. \ref{stable}(b).  We can observe the existence of incoherent oscillators in the edges of the coherent domain which is clearly illustrated with phase portraits of the oscillators in Fig. \ref{stable}(d).  Further increase in the number of perturbed oscillators leads to complete synchronization among the oscillators.  In Ref. [47], the authors have shown that the dependence of the lifetime of the amplitude chimera state with respect to the initial conditions becomes less important under the impact of noise.  In the case of local coupling, for fixed value of intensity of noise with different initial conditions, we can observe the lifetime of the amplitude chimera state remains unchanged.  Moreover, the lifetime of the amplitude chimera state decreases with an increase in the value of the intensity of noise for particular initial condition.
\begin{figure}[ht!]
\begin{center}
 \includegraphics[width=1.0\linewidth]{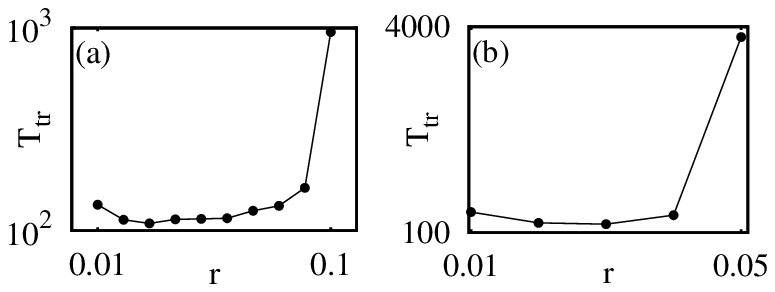}
\end{center}
\caption{Transient time of amplitude chimera state as a function of nonlocal coupling range $r=\frac{P}{N}$: (a) for $\varepsilon=5$, (b) $\varepsilon=8$ with $c=1.0$ and $N=100$.}
\label{nonlocal}
\end{figure}   
\subsection{Effect of system size}
\par Further we also analyze how the transient nature of the amplitude chimera state changes with respect to the system size and this is clearly illustrated with Figs. \ref{number}(a-c) for three different coupling interaction values.  From Fig. \ref{number}(a), we can observe that for an increase of system size $N$, initially the transient time $T_{tr}$ decreases and then it saturates at $T_{tr}=658$ for large $N=500$ with $\varepsilon=5$.  In the case for $\varepsilon=8$, for small values of $N$, $T_{tr}$ decreases and then it increases and finally it gets saturated at $T_{tr}=1673$ for $N=800$.  On the other hand, in the case of $\varepsilon=11$, we can observe that $T_{tr}$ decreases and it saturates at $T_{tr}=2954$ for $N=1000$.  Hence the saturation value of the number of oscillators in the system increases for an increase of coupling interaction.  It can also be noted that on increasing the strength of coupling interaction the saturated value of transient time $T_{tr}$ also increases. 
\subsection{Effect of nonlocal coupling}
 Also there arises a question as to whether transient time of amplitude chimera state is affected by the nonlocal coupling.  To answer this question we analyze the life-time of the amplitude chimera state with respect to nonlocal coupling range $r=\frac{P}{N}$ for $N=100$.  Here $P$ is the number of nearest neighbor and $P=1$ for local coupling.  This is illustrated in Figs. \ref{nonlocal}(a) and (b) for fixed nonisochronicity value $c=1$ (parameters are suitably chosen in the transient amplitude chimera state region).  For $\varepsilon=5$, we can observe that for an increase of $r$, the life time of amplitude chimera state increases and it becomes stable chimera (transients are left out upto $10^8$ time units) for $r=0.11$.  Similarly for increased value of coupling strength, $\varepsilon=8$, also the life time of amplitude chimera state increases and becomes stable for $r=0.06$.  It can be seen clearly that amplitude chimera state becomes stable for small coupling strength $\varepsilon$ with large coupling range and if the coupling strength is comparatively large compared to the former case it becomes stable even with small coupling range.   
\section{Conclusion}
\par In summary, we have investigated the existence of transient and stable amplitude chimera states and imperfect breathing chimera states in an array of locally coupled oscillators in one dimension under two different initial conditions.  The existence of imperfect breathing chimera state is confirmed through the localized set approach and by finding the relative phases of the oscillators.  The choice of cluster initial condition has been found to support the presence of the above mentioned dynamical states including amplitude chimera state and imperfectly breathing chimera state. We have also analyzed the stability nature of the amplitude chimera state under various circumstances such as variation of coupling strength, perturbation in initial state of the oscillators, change in the system size and system parameters.  In addition, we have found the stability of the transient and stable amplitude chimera states as well as traveling wave states for appropriate number of oscillators using Floquet theory.  We have found that the transient time of the amplitude chimera state increases if the system of oscillators are coupled through nonlocal interaction.
\section*{Acknowledgements}
 The work of KP and ML forms part of a research project sponsored by DST-SERB under Grant No. EMR/2014/001076.   ML also acknowledges the financial support under a NASI Senior Scientist Fellowship program and Council of Scientific and Industrial Research (CSIR), Government of India, research project under Grant No. 03(1331)/15 EMR-II.  The work of VKC is supported by the SERB-DST Fast Track scheme for young scientists under Grant No. YSS/2014/000175.  The work of MS forms part of a different research project sponsored by CSIR under Grant No. 03(1397)/17/EMR-II. 
\appendix
\section{CHARACTERISTIC MEASURE FOR STRENGTH OF INCOHERENCE}
\par Strength of incoherence \cite{21} is used to identify the nature of different dynamical states in the system, that will help us to detect interesting collective dynamical states such as synchronized state, desynchronized state, and the chimera state.  For this purpose we introduce a transformation $w_j=x_j-x_{j+1}$ \cite{21}, where $j=1,2,3,...,N$.  We divide the oscillators into $K$ bins of equal length $l=N/K$ and the local standard deviation $\sigma_l(m)$ is defined as   
\begin{eqnarray} 
\sigma_l(m) =\langle(\overline{ \frac{1}{l}\sum_{j=l(m-1)+1}^{ml} \vert w_j -\overline{w }\vert^2})^{1/2}\rangle_t,\nonumber\\
 m=1,2,...K.
\label{sig}
\end{eqnarray}
\par From this we can find the local standard deviation for every $K$ bins of oscillators that helps to find the strength of incoherence \cite{21} through the expression
\begin{equation} 
S =1-\frac{\sum_{m=1}^{K} s_m }{K},s_m =\Theta(\delta- \sigma_l(m) ),
\label{soi}
\end{equation}
where $\delta$ is the threshold value which is small. Here $\langle ... \rangle_t$ represents the average over time.  When $\sigma_l(m) $ is less than $\delta$, $s_m $ takes the value $1$, otherwise it is $0$.  Thus the strength of incoherence measures the amount of spatial incoherence present in the system which is zero for the spatially coherent  synchronized state.  It has the maximum value, that is $S =1$, for the completely incoherent desynchronized state and has intermediate values between 0 and 1 for chimera states and cluster states.  Further, to distinguish the amplitude chimera state and phase chimera state, we find the strength of incoherence in the amplitude domain $S_r$ as different from $S$.  For finding $S_r$, we use the same procedure as above with $w_{j}=r_{j}-r_{j+1}$ ($r_j^2=\sqrt{x_j^2+y_j^2}$).  Now $S_r$ can be used to clearly distinguish the phase and amplitude chimera state.  Since the amplitude of all the oscillators in the system are the same for phase chimera state and strength of incoherence in the amplitude domain is $S_r=0$ while $S$ varies between 0 and 1.  On the other hand both $S$ and $S_r$ have the values between 0 and 1 for amplitude chimera states.

\end{document}